\documentstyle[12pt]{article}

\newcommand{\om}{\omega}
\newcommand{\Om}{\Omega}  
\newcommand{\be}{\begin{equation}}
\newcommand{\ee}{\end{equation}}
\newcommand{\Dlt}{\Delta}
\newcommand{\dlt}{\delta}
\newcommand{\gm}{\gamma}
\newcommand{\vS}{{\bf S}}
\newcommand{\vB}{{\bf B}}
\newcommand{\al}{\alpha}
\newcommand{\bt}{\beta}
\newcommand{\vr}{{\bf r}}
\newcommand{\vn}{{\bf n}}
\newcommand{\ra}{\rightarrow}
\newcommand{\sgm}{\sigma}
\newcommand{\Gm}{\Gamma}

\begin{document}
\begin{center}
{\Large{\bf Superradiant Operation of Spin Masers} \\ [5mm]
V.I. Yukalov} \\ [3mm]

{\it Research Center for Optics and Photonics \\
Instituto de Fisica de S\~ao Carlos, Universidade de S\~ao Paulo \\
Caixa Postal 369, S\~ao Carlos, S\~ao Paulo 13560-970, Brazil \\ 

and \\

Bogolubov Laboratory of Theoretical Physics \\
Joint Institute for Nuclear Research, Dubna 141980, Russia}

\end{center}

\vskip 3cm

\begin{abstract}

The theory of spin superradiance, developed earlier for nuclear 
magnets, is generalized to a wider class of spin systems, such as 
granular magnets and molecular magnets. The latter may possess strong 
single-site magnetocrystalline anisotropy, whose role in nonlinear 
spin dynamics is analysed. Transient as well as pulsing superradiant 
regimes are described. These coherent regimes may be employed in the 
operation of spin masers.

\end{abstract}

\newpage

\section{Introduction}

Among different masers, generating radiation at microwave and radio
frequencies [1], there is a separate class of spin masers, in which 
the radiation process is due to moving spins. The role of a resonator
for such masers is played by a resonant electric circuit coupled with
a spin system. The peculiarity of spin masers is that their magnetodipole
radiation is rather weak and practically does not propagate into free 
space but mainly is taken up by a resonant coil surrounding the spin 
sample [2]. Nevertheless, because of a deep physical similarity with
other types of masers, radiating spin systems are called spin masers 
[3].

Resonant optical systems can display the effect of {\it superradiance}, 
which is a {\it coherent spontaneous emission}, when the radiation 
intensity is approximately proportional to the number of radiators
squared [4,5]. Spin systems may also exhibit this phenomenon [6] which, 
because of its direct analogy with optical {\it atomic superradiance},
is termed {\it spin superradiance}. The latter, in the same way as the
former, can be of two major types, transient and pulsing. The {\it
transient superradiance} occurs as a single sharp superradiant burst 
that may be accompanied by several small quickly diminishing oscillations. 
This regime happens when the system is prepared in an inverted strongly
nonequilibrium state, after which it is not influenced by any additional 
external fields. The {\it pulsing superradiance} corresponds to a long
train of superradiant bursts, which can be realized if the system is 
subject to a permanent pumping mechanism. Both types of superradiance 
were experimentally observed for several spin systems, including the 
transient [7--10] and pulsing [11--13] spin superradiance. The theory of 
spin superradiance was developed [14--20], being based on microscopic 
Hamiltonians and allowing for the first correct description of {\it 
purely self-organized superradiance}. Numerical simulations of spin 
superradiance, being a kind of computer experiments, were also realized 
[21--27].

These studies of spin superradiance were based on the Hamiltonian 
describing an ensemble of polarized nuclear spins interacting through
dipolar forces. Such a matter, for short, can be called a {\it nuclear
magnet}. The influence of hyperfine forces on nuclear spin superradiance
was also considered [28--30]. This attention to nuclear magnets has been
due to the fact that the experiments with spin superradiance [7--13]
had been accomplished for nuclear spins. In principle, this effect could
be realized for electronic spins as well, with a similar theoretical
description of electron spin superradiance [28,31,32].

In the present paper, the theory of spin superradiance is generalized 
to a wider class of materials, such as granular magnets and molecular 
magnets, which can possess higher spins and strong single-site magnetic
anisotropy. The theory is also improved by taking a more accurate account
of retardation effects.

\section{Materials Characteristics}

The experiments on spin superradiance till now have been accomplished
with nuclear magnets. The first observation of transient spin 
superradiance [7,8] and its confirmation [9] were done for propanediol
C$_3$H$_8$O$_2$. An active substance here is the ensemble of proton
spins with the density $\rho_H\sim 4\times 10^{22}$ cm$^{-3}$. The spins
were polarized, by means of dynamic nuclear polarization, parallel to an 
external magnetic field $B_0\sim 1$ T, which corresponds to the Zeeman
frequency $\om_0\sim 10^8$ Hz. The material was kept at low temperature
$T\sim 0.1$ K, which resulted in strong suppression of the nuclear 
spin-lattice relaxation. The corresponding longitudinal relaxation time 
was $T_1\sim 10^5$ s. The transverse dephasing time, due to dipolar 
interactions, was $T_2\sim 10^{-5}$ s. The sample was coupled to a 
resonant electric circuit with a quality factor $Q\sim 100$ and a ringing
time $\tau\sim 10^{-6}$ s.

In other experiments on transient spin superradiance [10] butanol
C$_4$H$_9$OH and ammonia NH$_3$ were used. These materials are rich with 
protons of density $\rho_H\sim 10^{23}$ cm$^{-3}$. An external magnetic 
field $B_0\sim 1$ T defined the Zeeman frequency $\om_0\sim 10^8$ Hz.
The experiments were carried out at low temperature, resulting in the 
long spin-lattice relaxation time $T_1\sim 4\times 10^4$ s up to $10^5$ 
s. The transverse dephasing time was $T_2\sim 10^{-5}$ s. The quality 
factor of the resonant electric circuit was $Q\sim 30$, and the ringing 
time $\tau\sim 5\times 10^{-7}$ s.

The experiments on pulsing spin superradiance [11--13] employed the ruby
crystal Al$_2$O$_3$. The active nuclei here are $^{27}$Al, with spin
$I=5/2$ and density $\rho_{Al}\sim 4\times 10^{22}$ cm$^{-3}$. The
crystal was oriented in an external magnetic field $B_0\sim 1$T so that
a fully resolved structure of its five $\Dlt m=\pm 1$ transitions could 
be observed. Then, if a resonant circuit is tuned to a selected transition
line, $^{27}$Al spins form an {\it effective two-level system}. In 
experiments, the circuit was tuned to the central 
$\{-\frac{1}{2},\frac{1}{2}\}$ line, with a transition frequency 
$\om_0\sim 10^8$ Hz. At low temperatures $T\sim 1$ K, the spin-lattice
relaxation time was $T_1\sim 10^5$ s. The transverse relaxation time
was $T_2\sim 10^{-5}$ s. The quality factor of the resonant circuit 
was $Q\sim 100$, with the ringing time $\tau\sim 10^{-6}$ s. The 
inversion of spin polarization was permanently supported by means of 
dynamic nuclear polarization with the pumping rate $\gm_1^*=10$ s.

Wishing to extend the possibility of realizing spin superradiance 
for other types of spin systems, the first such materials that come 
to mind are {\it granular magnets}. These are composed of magnetic 
nanoparticles of diameters between $10\;\AA$ to $10^4\;\AA$. Each 
nanoparticle is a superparamegnetic cluster of an effective spin $S$ 
that can be sufficiently high. There exists a large variety of magnetic
nanoparticles [33,34] which can be formed by simple metals, such as
Ni, Fe, Co, and Hg or their oxides, as NiO and Fe$_2$O$_3$. Many magnetic
nanoparticles are made of different alloys, such as NiFe$_2$O$_4$,
Nd$_2$Fe$_{14}$B, Pr$_2$Fe$_{14}$B, Tb$_2$Fe$_{14}$B, Dy$_2$Fe$_{14}$B,
Pr$_2$Co$_{14}$B, Sm$_1$Fe$_{11}$Ti$_1$, Sm$_1$Fe$_{10}$V$_2$,
Sm$_2$Fe$_{17}$N$_{23}$, Sm$_2$Fe$_{17}$C$_{22}$, Sm$_2$Co$_{17}$, and
SmCo$_5$. An ensemble of magnetic nanoparticles could be polarized in 
an external magnetic field, after which, inverting the latter, one
would get an inverted nonequilibrium system. The following process 
should be similar to that developing in an ensemble of nuclear spins.
The main disadvantage of granular magnets is that composing them
nanoparticles vary in size and shape. It is practically hardly possible
to make a system of nanoparticles being almost identical. And an essential
variation of the properties of radiating objects leads to a large 
inhomogeneous broadening, which hinders the feasibility of achieving 
a good level of coherence.

Another class of composite objects, possessing nonzero spin, are magnetic
molecules [35]. These molecules can form crystalline materials where all 
magnetic clusters are well defined with the same shape, size, and
orientation, because of which the inhomogeneous broadening has to be
very low. Such materials, composed of magnetic molecules are termed
{\it molecular magnets}. There are many different magnetic molecules
[36,37] having in their ground state nonzero total spins $S$. For example,
the dodecanuclear manganese cluster 
[Mn$_{12}$O$_{12}$(CH$_3$COO)$_{16}$(H$_2$O)$_4$]$\cdot$2CH$_3$COOH$\cdot$4H$_2$O
is a molecule with spin $S=10$. It has a rather strong single-site 
anisotropy characterized by a parameter $D\approx 0.7$ K, which makes 
the anisotropy barrier $DS^2\approx 70$ K. At low temperatures, lower 
than the blocking temperature $T_B\approx 3$ K, the magnetization of a 
molecular crystal is preserved during the relaxation time $\approx 10^7$ 
s. The size of a molecule is about $10\;\AA$ and the distance between 
neighbouring molecules in the molecular crystal is about $14\;\AA$. The
interaction between molecules is through dipolar forces, with an energy
$\approx 0.1$ K or $10^{10}$ Hz. More information on the properties of
this molecule can be found in Refs. [38--48].

The octanuclear iron molecular cluster 
[Fe$_8$O$_2$(OH)$_{12}$(tacn)$_6$]$^{8+}$, where "tacn" stands for 
the organic ligand triazacyclononane, also has a high spin $S=10$. Its 
magnetic anisotropy is $D\approx 0.3$ K, hence the anisotropy barrier 
is $DS^2\approx 30$ K. The blocking temperature is 
$T_B\approx 1$ K, below which the relaxation time for a molecular crystal 
is about $10^5$ s. The size of a molecule and the intermolecular distance 
in a crystal are close to those for the molecule Mn$_{12}$ mentioned 
above. Other characteristics of the molecule can be found in Refs. 
[36,37,49--53].

The molecule [Cr(CNMnL)$_6$](ClO$_4$)$_9$, where L is a neutral 
pentadentate ligand, has in its ground state the total spin $S=27/2$
(see [37]). The molecule [(PhSiO$_2$)$_6$Cu$_6$(O$_2$SiPh)$_6$] 
possesses the spin $S=3$ (see [54]). And the molecule 
K$_6$[V$_{15}^{4+}$As$_6$O$_{42}$(H$_2$O)]$\cdot$8H$_2$O has the spin 
$S=1/2$ and no magnetic anisotropy [55,56]. There is a number of other
molecules [37] with nonzero total ground-state spin. Some molecules may
have zero spin in their ground state but a finite spin in excited states
[57--59].

The relaxation of the total magnetization of a crystal, consisting of
many magnetic molecules, occurs because of the axial degeneracy of spin
direction in each molecule. At zero external magnetic field, the spin of
a single molecule can be directed either up or down, so that the
equilibrium state of an ensemble of these molecules corresponds to
zero total magnetization. At temperatures higher than the blocking
temperature $T_B$, the relaxation is rather fast and is due to thermal
fluctuations. At low temperatures below $T_B$, the relaxation is very
slow, so that the polarization of a molecular magnet can be blocked for
months. At such low temperatures, the relaxation is characterized as 
quantum spin tunneling between the degenerate states. To understand the
low-temperature behaviour of magnetic molecules, it is necessary to
take into account both the internal effects, caused by atoms composing
each molecule [60--62], as well as external interactions between the
molecules forming a crystal [63--65]. Taking account of dipole 
interactions between molecules is crucial for correctly describing the
relaxation in a molecular magnet [42,51,52,56,66].

In this way, there exists a large variety of different objects, such as 
nuclei, granules, and molecules, each of which can be considered as an 
entity, like a particle with an effective spin that can vary in a wide 
diapason. The main interactions between such magnetic particles, forming 
a solid, are dipole interactions. A sample, composed of these particles, 
can be polarized and at low temperatures the polarization can be preserved
for a very long time. A specific feature of magnetic particles with high 
spin is the presence of the single-site magnetic anisotropy, which one 
has to take into account when considering collective spin dynamics. The 
problem, to be addressed in the following sections, is how coherent spin 
radiation can arise is such materials made of objects with high spins and 
magnetic anisotropy. A special attention will be paid to the possibility 
of realizing spin superradiance that is a self-organized process 
developing without imposing on the spin system an initial transverse 
coherence.

\section{Types of Coherence}

Before  passing to the development of a generalized theory of spin
superradiance, a few words are to be said for concretizing the term 
"coherence" that will be repeatedly used in what follows. Generally,
one uses this term in two different meanings. One widespread usage 
implies under a coherent state of a many-particle system just a pure
quantum state characterized by the same wave function for all particles.
Then an incoherent state is a mixed state described by a density matrix 
[67]. In the theory of nuclear magnetic resonance [68,69], coherence 
usually means the existence of transverse magnetization and, more 
generally, the existence of non-diagonal matrix elements. One also tells
[68,69] that the existence of the transverse magnetization means phase
coherence, as opposed to amplitude coherence associated with a nonzero
longitudinal polarization. To formalize these definitions for a system
of $N$ spins, let us introduce the notation
$$
S^z \equiv \frac{1}{N}\; \sum_{i=1}^N S_i^z \; , \qquad
S^\pm \equiv \frac{1}{N} \; \sum_{i=1}^N S_i^\pm \; ,
$$
where $S_i^z$ is the $z$-component of the spin operator and $S_i^\pm$ 
are the raising and, respectively, lowering spin operators. Denote the
statistical averaging by the angle brackets $<\ldots>$. Then a nonzero
$<S^z>\neq 0$ means {\it longitudinal coherence}, or {\it diagonal
coherence}, or {\it amplitude coherence}. While a nonzero $<S^\pm>\neq 0$
signifies the existence of {\it transverse coherence}, or {\it nondiagonal
coherence}, or {\it phase coherence}.

Another possibility could be to tell that a nonzero $<S^z>\neq 0$ 
corresponds to {\it state coherence}. When the whole system is in pure 
state, then $<S^z>=\pm S$, which can be named {\it pure coherent state},
to distinguish it from the case of {\it partially coherent state}, when
$0<|<S^z>|<S$. This definition of state coherence is in agreement with
the quantum-mechanical understanding of a coherent state as of a pure
state. It also agrees with the definition of coherent states in quantum
field theory as of eigenvalues of field operators [70]. In many 
applications, the operator $S^z$ characterizes a population difference. 
Hence, nonzero $<S^z>$ may be associated with {\it population coherence}.
Being related to the determination of eigenfunctions of operators, the 
state coherence may also be called {\it quantum coherence}.

Since the raising and lowering operators describe transitions between 
quantum states, the existence of nonzero $<S^\pm>\neq 0$ can be ascribed 
to {\it transition coherence}. This type of coherence is closely connected
with the existence of coherent radiation, because of which it may be 
termed {\it radiation coherence}. Also, this recalls the classical 
understanding of coherence as of synchronous motion of several objects, 
which makes it sometimes possible to use the name of {\it classical
coherence}. Synchronous motion is often termed as the motion in phase.
That is why the term of {\it phase coherence} is appropriate here.

These types of coherence are not strictly correlated with each other. 
A spin system may possess one of them or both, or neither. But often 
they are complimentary to each other, as it happens in the process of
superradiance. Then, at the initial time, the spin system has to be 
prepared in a well polarized state, thus, displaying state coherence.
In addition, this state must be strongly nonequilibrium. Transition 
coherence, at the initial time, should be absent. It has to develop 
in a self-organized way owing to mutual spin correlations, which can 
be realized trough a feedback field. Usually, the maximal transition 
coherence in superradiance develops at the moment when the state 
coherence is minimal.

\section{Spin Hamiltonian}

Consider an ensemble of $N$ spins $\vS_i$ enumerated by the index 
$i=1,2,\ldots,N$. These spins can correspond either to nuclei, or to
granules, or to molecules. The Hamiltonian of the system,
\be
\label{1}
\hat H = \sum_i \hat H_i + \frac{1}{2}\; 
\sum_{i\neq j} \hat H_{ij} \; ,
\ee
contains the terms $\hat H_i$, related to individual spins, and the
terms $\hat H_{ij}$ describing spin interactions. An individual term
\be
\label{2}
\hat H_i = -\mu_0\vB\cdot\vS_i - D(S_i^z)^2
\ee
includes the Zeeman energy and a part characterizing magnetic anizotropy.
In the case of nuclei, $\mu_0= g_S\mu_N =\hbar\gm_S$, where $g_S$ is
the Land\'e factor; $\mu_N$, nuclear magneton; $\gm_S$, gyromagnetic
ratio. For nuclei, $\mu_0$ can be positive as well as negative. When 
$S$ is an electron spin, $\mu_0=-g_S\mu_B=-\hbar\gm_S$, where $\mu_B$
is the Bohr magneton. Then $\mu_0$ is negative. The anisotropy term, 
with an anisotropy constant $D$, is nontrivial only for spins higher
than $1/2$. Positive $D>0$ implies an easy-axis anisotropy, while
$D<0$ means an easy-plane anisotropy. The pair terms in the Hamiltonian
(1) describe dipolar interactions
\be
\label{3}
\hat H_{ij} = \sum_{\al\bt} C_{ij}^{\al\bt} S_i^\al S_j^\bt
\ee
through the dipolar tensor
\be
\label{4}
C_{ij}^{\al\bt} = \frac{\mu_0^2}{r_{ij}^3} \left ( \dlt_{\al\bt} - 
3 n_{ij}^\al n_{ij}^\bt\right ) \; ,
\ee
in which $\al,\bt = x,y,z$ and
$$
r_{ij} \equiv |\vr_{ij}| \; , \vn_{ij}\equiv \frac{\vr_{ij}}{r_{ij}}\; ,
\qquad \vr_{ij} \equiv \vr_i - \vr_j \; .
$$
The dipolar tensor enjoys the properties
\be
\label{5}
\sum_\al C_{ij}^{\al\al} = 0 \; , \qquad 
\sum_{j(\neq i)} C_{ij}^{\al\bt} = 0 \; ,
\ee
of which the first is exact and the second is asymptotically exact
for a macroscopic sample with a large number of spins $N\gg 1$. The
total magnetic field
\be
\label{6}
\vB = B_0{\bf e}_z + (B_1 + H){\bf e}_x
\ee
consists of an external longitudinal magnetic field $B_0$, transverse
magnetic field $B_1$, and a feedback field $H$ of the resonant electric
circuit. In what follows, the longitudinal magnetic field is assumed to 
be constant and directed so that
\be
\label{7}
\mu_0 B_0 < 0 \; .
\ee
In particular, for electronic spins, $B_0>0$, since $\mu_0$ is negative.

For what follows, it is convenient to pass to raising and lowering operators
$$
S_i^\pm \equiv S_i^x \pm i S_i^y \; ,
$$
which are Hermitian conjugated with each other. Also, introduce the notation
$$
a_{ij} \equiv C_{ij}^{zz} \; , \qquad b_{ij} \equiv
\frac{1}{2}\left ( C_{ij}^{xz} + i C_{ij}^{yz}\right ) \; ,
$$
\be
\label{8}
c_{ij} \equiv \frac{1}{4}\left ( C_{ij}^{xx} + 2i C_{ij}^{xy} -
C_{ij}^{yy}\right ) \; .
\ee
Then the individual term (2) can be written as
\be
\label{9}
\hat H_i = -\mu_0 B_0 S_i^z - D(S_i^z)^2 -\;
\frac{1}{2}\; \mu_0 (B_1 + H) \left ( S_i^+ + S_i^-\right ) \; .
\ee
And the interaction term (3) takes the form
$$
\hat H_{ij} = a_{ij}\left ( S_i^z S_j^z -\; \frac{1}{2}\; S_i^+ S_j^-
\right ) + 2b_{ij}^* S_i^+ S_j^z + 2b_{ij} S_i^- S_j^z +
$$
\be
\label{10}
+ c_{ij}^* S_i^+ S_j^+ + c_{ij} S_i^- S_j^- \; .
\ee
Thus, the spin system is described by the Hamiltonian (1), with 
the terms (9) and (10).

\section{Electric Circuit}

The external fields $B_0$ and $B_1$ are assumed to be given. What 
is not yet defined is the resonator feedback field $H$, produced by 
a resonant electric circuit. Let the circuit be characterized by 
resistance $R$, inductance $L$, and capacity $C$. The spin sample
is inserted into a coil of $n$ terns, length $l$ and cross-section
area $A_c$. The electric current in the circuit is determined by the
Kirchhoff equation
\be
\label{11}
L\; \frac{dj}{dt} + Rj + \frac{1}{C} \int_0^t j(t')\; dt'=
\tilde E - \; \frac{d\Phi}{dt} \; ,
\ee
in which $\tilde E$ is an electromotive force and $\Phi$ is a magnetic
flux
\be
\label{12}
\Phi = \frac{4\pi}{c}\; n A_c\eta M_x 
\ee
formed by the $x$-component of the magnetization density
\be
\label{13}
M_x = \frac{\mu_0}{V}\; \sum_i < S_i^x > \; ,
\ee
with the brackets $<\ldots>$ implying statistical averaging. The filling 
factor $\eta$ is approximately equal to $\eta\approx V/V_c$, where $V$ 
is the sample volume and $V_c\equiv A_c l$ is the coil volume.

The current, circulating over the coil, produces a magnetic field
\be
\label{14}
H = \frac{4\pi n}{cl}\; j \; ,
\ee
with $c$ being the light velocity. Hence, Eq. (11) may be rewritten for
the field (14). Introduce the circuit natural frequency
\be
\label{15} 
\om \equiv \frac{1}{\sqrt{LC}} \qquad \left ( L \equiv 
\frac{4\pi n^2 A_c}{c^2 l} \right )
\ee
and the circuit ringing time
\be
\label{16}
\tau \equiv \frac{1}{\gm} \qquad \left ( \gm \equiv \frac{R}{2L}
\right ) \; ,
\ee
with the related circuit damping
\be
\label{17}
\gm = \frac{\om}{2Q} \qquad \left ( Q \equiv \frac{\om L}{R}
\right ) \; ,
\ee
where $Q$ is the quality factor. Define the reduced electromotive force
\be
\label{18}
h \equiv \frac{c\tilde E}{n A_c\gm} \; .
\ee
Then the Kirchhoff equation (11) can be transformed to the equation
\be
\label{19}
\frac{dH}{dt} + 2\gm H + \om^2 \int_0^t H(t')\; dt'= 
\gm h - 4\pi\eta\; \frac{dM_x}{dt}
\ee
for the feedback magnetic field produced by the coil.

The feedback equation (19) can be presented in another equivalent form
[15--18] which is very useful for solving the evolution equations. For
this purpose, we apply to Eq. (19) the method of Laplace transforms
and use the transfer function
$$
G(t) = \left ( \cos\tilde\om t -\;
\frac{\gm}{\tilde\om}\; \sin\tilde\om t\right ) e^{-\gm t} \; ,
$$
with $\tilde\om\equiv\sqrt{\om^2 -\gm^2}$. Then  from the feedback 
equation (19) we obtain
\be
\label{20}
H = \int_0^t G(t-t') \left [ \gm h(t') - 4\pi\eta \dot{M}_x(t')
\right ] \; dt'\; ,
\ee
where $\dot{M}_x \equiv dM_x/dt$.

\section{Evolution Equations}

First, we write the Heisenberg equations
$$
i\hbar \; \frac{dS_i^\al}{dt} = \left [ S_i^\al,\hat H\right ]
$$
for the spin operators, using the commutation relations
$$
[S_i^-,\; S_j^+] = - 2\dlt_{ij} S_i^z \; , \qquad [S_i^z,\; S_j^\pm] =
\pm\dlt_{ij} S_i^\pm \; , \qquad
\left [ S_i^-,\;(S_j^z)^2\right ] = \dlt_{ij}\left ( S_i^- S_i^z + 
S_i^z S_i^- \right ) \; .
$$
For the lowering operator, this gives
$$
i\hbar\; \frac{dS_i^-}{dt} = -\mu_0 B_0 S_i^- + \mu_0 (B_1 + H) S_i^z -
D(S_i^- S_i^z + S_i^z S_i^-) +
$$
\be
\label{21}
+ \sum_{j(\neq i)} \left [ a_{ij}\left ( S_i^- S_j^z + \frac{1}{2}\;
S_i^z S_j^-\right ) + b_{ij} S_i^- S_j^- + 
b_{ij}^*\left ( S_i^- S_j^+ - 2 S_i^z S_j^z \right ) -
2c_{ij}^* S_i^z S_j^+ \right ] \; .
\ee
The equation for the raising operator is obtained from Eq. (21) by 
Hermitian conjugation. The equation for $S_i^z$ is
$$
i\hbar\; \frac{dS_i^z}{dt} = \frac{1}{2}\; \mu_0 (B_1 + H) 
\left ( S_i^- - S_i^+\right ) +
$$
\be
\label{22}
+ \sum_{j(\neq i)}\left [ \frac{1}{4}\; a_{ij}\left ( S_i^- S_j^+ -
S_i^+ S_j^-\right ) + b_{ij}^* S_i^+ S_j^z - b_{ij} S_i^- S_j^z +
c_{ij}^* S_i^+ S_j^+ - c_{ij}  S_i^- S_j^-\right ] \; .
\ee

To analyze these equations, we employ the scale separation approach 
[14,17,18,71]. Notice, first, that in Eqs. (21) and (22), we may
separate the terms
$$
\xi_0 \equiv \frac{1}{\hbar}\; \sum_{j(\neq i)} \left ( a_{ij} S_j^z 
+ b_{ij}^* S_j^+ + b_{ij} S_j^-\right ) \; ,
$$
\be
\label{23}
\xi \equiv -\; \frac{i}{\hbar} \sum_{j(\neq i)} \left ( 
\frac{1}{2}\; a_{ij} S_j^- -  
2b_{ij}^* S_j^z - 2c_{ij}^* S_j^+ \right ) \; ,
\ee
which play the role of local fields acting on spins. Statistical 
averages of these fields, with the usage of the uniform approximation, 
are zero owing to the equalities
$$
\sum_{j(\neq i)} a_{ij} = \sum_{j(\neq i)} b_{ij} = 
\sum_{j(\neq i)} c_{ij} = 0
$$
following from the properties (5). At the same time, statistical 
averages of the spin operators are certainly nonzero. Therefore the 
local fields (23), acting on a short scale, can be treated as operator 
variables of nature different from the spin operators. Such local 
fields may be modelled by random variables [68,69,72]. Thus, we have two
types of variables in the system, spin operators $S_i^-$, $S_i^+$,
$S_i^z$, and stochastic fields $\xi_0,\; \xi,\; \xi^*$. The former are
responsible for long-range global phenomena while the latter, for
short-range local effects. The stochastic fields describe fast 
fluctuations in the local surrounding of each spin. The existence of 
such fluctuations yields inhomogeneous {\it dynamic broadening}.

To make the problem closed, it is necessary to define the stochastic
averages over the random fields (23). The latter can be treated as white 
noise with the stochastic averages
$$
\ll \xi_0(t)\gg \; = \; \ll \xi(t)\gg \; = 0 \; , \qquad
\ll \xi_0(t)\xi_0(t')\gg \; = 2\gm_3\dlt(t-t') \; ,
$$
\be
\label{24}
\ll \xi_0(t)\xi(t')\gg \; = \; \ll \xi(t)\xi(t')\gg \; = 0 \; ,
\qquad \ll \xi^*(t)\xi(t')\gg \; = 2\gm_3\dlt(t-t') \; ,
\ee
where $\gm_3$ is the width of dynamic broadening. The method of modelling 
local fields by random variables can be called {\it randomization of 
local fields}.

Averaging over spin operators, which are responsible for long-range 
phenomena, we may employ the mean-field approximation
\be
\label{25}
<S_i^\al S_j^\bt>\; = \; < S_i^\al><S_j^\bt> \qquad (i\neq j) \; .
\ee
This kind of approximation can be used only for the pairs of spins at
different sites $i\neq j$, since $S_i^\al$ and $S_j^\bt$ commute in
such a case. But the spin operators do not commute for $i=j$. Hence
the averages $<S_i^\al S_i^\bt>$ cannot be factorized as above. Notice
that for spin $S=1/2$, the anisotropy term in the Hamiltonian does not
contribute to the equations of motion, since $(S_i^z)^2=1/4$. The problem
of spin decoupling, more general than condition (25), has been considered 
by several authors [73--76]. The term in Eq. (21), caused by the magnetic 
anisotropy, can be decoupled as follows:
\be
\label{26}
<S_i^- S_i^z + S_i^z S_i^-> \; =\left ( 2 -\; \frac{1}{S}\right )
<S_i^-><S_i^z> \; .
\ee
This presentation enjoys correct limiting properties. For $S=1/2$, it
nullifies; while for $S\ra\infty$, when spins behave classically, a 
simple factorization occurs. The decouplings (25) and (26) do not take
account of spin-spin correlations, which can be incorporated into the
evolution equations by including the term describing spin attenuation, 
with the related spin-spin relaxation width $\gm_2$. To allow for 
the influence of lattice, account must be taken of the spin-lattice 
relaxation, with the corresponding relaxation parameter $\gm_1$.

Let us average the equations of motion (21) and (22) over the spin 
operators, not touching the stochastic fields (23). In so doing, we 
introduce the following definitions. The variable
\be
\label{27}
x\equiv \frac{1}{S}\; < S_i^->
\ee
describes the rotation of transverse spin components. As is discussed 
in Section 3, this average is connected with the arising transition 
coherence, or transverse coherence, or radiation coherence. The degree 
of such a coherence can be characterized by the real function
\be
\label{28}
y \equiv \frac{1}{S^2}\; < S_i^+><S_i^-> \; = |x|^2 \; .
\ee
And the longitudinal spin polarization is given by
\be
\label{29}
z\equiv \frac{1}{S}\; <S_i^z> \; .
\ee
Recall that the radiation of spins happens at radio-frequencies whose 
wavelengths are much larger than the mean distance between spins. 
Therefore, it is admissible to employ the uniform approximation, assuming
that the functions (27) to (29) do not depend on spatial variables.

Note that, instead of resorting to the uniform approximation for the
functions (27) to (29), it would be possible to work with the following
arithmetical averages: {\it transition function}
$$ 
x \equiv \frac{1}{NS}\; \sum_{i=1}^N < S_i^-> \; ,
$$
{\it coherence intensity}
$$
y \equiv \frac{1}{N^2S^2}\; \sum_{i\neq j}^N < S_i^+ S_j^-> \; ,
$$
and longitudinal polarization, or {\it spin polarization}
$$
z\equiv \frac{1}{NS}\; \sum_{i=1}^N < S_i^z> \; .
$$
The resulting equations would be absolutely the same.

Le us define the Zeeman frequency
\be
\label{30}
\om_0 \equiv \frac{1}{\hbar} |\mu_0 B_0| \; ,
\ee
the effective transition frequency
\be
\label{31}
\tilde\om_0 \equiv \om_0 - (2S -1 ) Dz \; ,
\ee
and introduce the notation
\be
\label{32}
f \equiv - \; \frac{i}{\hbar}\; \mu_0 (B_1 + H) + \xi
\ee
for an effective force acting on spins. Then from Eqs. (21) and (22), 
we obtain the evolution equations for the functional variables (27) 
to (29):
\be
\label{33}
\frac{dx}{dt} = - i \left ( \tilde\om_0 +\xi_0 - i\gm_2\right ) x
+ f z \; ,
\ee
\be
\label{34}
\frac{dy}{dt} = - 2\gm_2 y \left ( x^* f + f^* x\right ) z \; ,
\ee
\be
\label{35}
\frac{dz}{dt} = - \; \frac{1}{2} \left ( x^* f + f^* x \right ) -
\gm_1( z -\sgm) \; .
\ee
These equations are assumed to be complimented by the initial conditions
$$
x(0) = x_0 \; , \qquad y(0) = y_0 \; , \qquad z(0)=z_0 \; .
$$
Equations (33) to (35) are stochastic differential equations, since they
contain the stochastic variables $\xi_0$ and $\xi$. Getting stochastic 
equations is the price for making them closed. These also are nonlinear 
equations because of the resonator feedback field entering through the 
effective force (32). This resonator field is given by Eq. (20), where
one has to substitute the magnetization density
$$
M_x = \frac{1}{2}\; \rho\mu_0 S ( x^* + x) \qquad
\left ( \rho \equiv \frac{N}{V}\right ) \; .
$$
Due to the integral form (20), Eqs. (33) to (35) plus (20) compose a 
system of stochastic nonlinear integro-differential equations.

\section{Stochastic Averaging}

The system of equations (33) to (35) plus (20) looks rather complicated.
Nevertheless, it can be essentially simplified by invoking the {\it 
method of stochastic averaging} [17,18,71], which is a generalization
of multiscale averaging techniques to stochastic differential equations.
The applicability of the method is due to the existence of several small
parameters.

The spin-lattice and spin-spin relaxation parameters are assumed to be 
small as compared to the Zeeman frequency,
\be
\label{36}
\frac{\gm_1}{\om_0} \ll 1 \; , \qquad \frac{\gm_2}{\om_0} \ll 1 \; .
\ee
The interaction energy of spins with the resonator field is proportional
to the {\it natural width}
\be
\label{37}
\gm_0 \equiv \frac{\pi}{\hbar}\; \eta\rho\mu_0^2 S \; .
\ee
Since $\gm_2\sim n_0\rho\mu_0^2 S^2/\hbar$, where $n_0$ is the number of
nearest neighbors, then
$$
\gm_0 \sim \; \frac{\pi\eta}{n_0 S}\; \gm_2 < \gm_2 \; .
$$
Hence, the natural width (37) is small, as well as the width of dynamic
broadening,
\be
\label{38}
\frac{\gm_0}{\om_0} \ll 1 \; , \qquad \frac{\gm_3}{\om_0} \ll 1 \; .
\ee

Here we do not explicitly consider hyperfine interactions, 
surmising that their influence can be included in the values of 
the related relaxation parameters. In order to estimate the impact 
of these interactions, e.g. on the magnitude of the dynamic broadening, 
it is necessary to discriminate the cases when the radiating objects 
are nuclear spins or electronic spins. Recall that the spins of 
magnetic molecules are of electronic nature. The dynamic broadening 
of electronic spins, caused by dipolar and hyperfine interactions, 
respectively, is $\gm_3\sim \rho_e\mu_e^2$ and 
$\gm_3'\sim\rho_n\mu_n\mu_e$, where $\rho_e$ and $\rho_n$ are the 
densities of electronic or nuclear spins, with $\mu_e\equiv g_S\mu_BS$ 
and $\mu_n\equiv g_I\mu_NI$ being the electronic and nuclear moments. 
Similarly, denoting by capital letters the relaxation parameters for 
nuclear spins, we have the dynamic broadening, due to dipolar or 
hyperfine interactions as $\Gm_3\sim\rho_n\mu_n^2$ and 
$\Gm_3'\sim\rho_e\mu_e\mu_n$. From here, the following relations are 
valid:
$$
\frac{\gm_3'}{\gm_3} \sim \frac{\rho_n\mu_n}{\rho_e\mu_e} \; , \qquad
\frac{\Gm_3'}{\Gm_3} \sim \frac{\rho_e\mu_e}{\rho_n\mu_n} \; .
$$
Taking into account the equalities
$$
\frac{\mu_e}{\mu_n} = \frac{g_S\mu_BS}{g_I\mu_NI} \; , \qquad
\frac{\mu_B}{\mu_N} = \frac{m_p}{m_e} = 1836 \; ,
$$
we see that the hyperfine interactions do not play an important role for
electronic spins but can be rather important in the case of nuclear
spins [28--30].

The external transverse magnetic field is taken in the form
\be
\label{39}
B_1 = h_0 + h_1\cos\om t \; .
\ee
And let the resonant part of the reduced electromotive force (18) be
presented as
\be
\label{40}
h= h_c\cos\om t \; .
\ee
Introduce the notation
\be
\label{41}
\nu_0 \equiv \frac{\mu_0h_0}{\hbar} \; , \qquad 
\nu_1 \equiv \frac{\mu_0h_1}{2\hbar} \; , \qquad 
\nu_c \equiv \frac{\mu_0h_c}{2\hbar} \; .
\ee
The amplitudes of the fields (39) and (40) are supposed to be small, 
in the sense that
\be
\label{42}
\frac{|\nu_0|}{\om_0} \ll 1 \; , \qquad 
\frac{|\nu_1|}{\om_0} \ll 1 \; , \qquad 
\frac{|\nu_c|}{\om_0} \ll 1 \; .
\ee

The influence of stochastic fields has to be considered as weak, since
the stochastic averages (24) are proportional to the dynamic broadening
width that, according to Eq. (38), is small. In this way, the effective
force (32) can be treated as weak.

Finally, the resonant circuit is assumed to be of good quality, implying 
that
\be
\label{43}
\frac{\gm}{\om} \ll 1 \qquad (Q\gg 1) \; .
\ee

The existence of the listed small parameters shows that the transition 
function $x$, defined by Eq. (33), is to be considered as fast, compared 
to the slow functions $y$ and $z$, satisfying Eqs. (34) and (35).
Conversely, $y$ and $z$ are {\it temporal quasi-invariants} with respect
to $x$.

The resonator field $H$, in the first approximation, may be found 
by iterating Eq. (20) with the solution of Eq. (33) of zero order 
with respect to small parameters, that is with 
$x\simeq x_0\exp(-i\tilde\om_0t)$, where $z$ is a quasi-invariant.
This iteration yields
\be
\label{44}
\frac{\mu_0H}{\hbar} = i ( \al x - \al^* x^*) + 2\bt\cos\om t \; ,
\ee
where $\al$ is the coupling function of spins with the resonator 
feedback field,
\be
\label{45}
\al = \gm_0 \tilde\om_0\left [ 
\frac{1-\exp\{-i(\om-\tilde\om_0)t-\gm t\}}{\gm+i(\om-\tilde\om_0)} +
\frac{1-\exp\{i(\om+\tilde\om_0)t-\gm t\}}{\gm-i(\om+\tilde\om_0)} 
\right ] \; ,
\ee
and $\bt$ is the coupling function of spins with the electromotive 
force,
\be
\label{46}
\bt = \frac{\nu_c}{2}\left ( 1  - e^{-\gm t}\right ) \; .
\ee
Clearly, the action of the feedback field can be efficient only in the
case of a resonant coupling, which requires the {\it resonance condition}
\be
\label{47}
\frac{|\tilde\Dlt|}{\om} \ll 1 \; , \qquad 
\tilde\Dlt\equiv \om - |\tilde\om_0| \; .
\ee
Note that the effective frequency (31) can, in general, be positive as 
well as negative. The spin-feedback coupling (45) simplifies if the 
resonance is good, which means that $|\tilde\Dlt|<\gm$. In such a case,
Eq. (45) acquires the simple form
\be
\label{48}
\al \approx \frac{\gm\gm_0\tilde\om_0}{\gm^2+\tilde\Dlt^2}
\left ( 1  - e^{-\gm t}\right ) \; .
\ee
More generally, the coupling function (45) is complex, with its real 
and imaginary parts being
$$
{\rm Re}\;\al = \frac{\gm\gm_0\tilde\om_0}{\gm^2+\tilde\Dlt^2}\left [
1 - \left ( \cos\tilde\Dlt t - \; \frac{\tilde\Dlt}{\gm} \; 
\sin\tilde\Dlt t\right ) e^{-\gm t} \right ] \; ,
$$
$$
{\rm Im}\; \al = -\; \frac{\gm\gm_0|\tilde\om_0|}{\gm^2+\tilde\Dlt^2}
\left [ \frac{\tilde\Dlt}{\gm} \;  - \left ( \sin\tilde\Dlt t + 
\frac{\tilde\Dlt}{\gm} \; \cos\tilde\Dlt t\right ) e^{-\gm t} \right ]\; ,
$$
where the resonance condition (47) is used.

Expression (44) is to be substituted into Eqs. (33) to (35). With this 
in mind, we define the {\it collective frequency}
\be
\label{49}
\Om \equiv \tilde\om_o - z{\rm Im}\; \al 
\ee
and the {\it collective attenuation}
\be
\label{50}
\Gm \equiv \gm_2 -  z{\rm Re}\; \al \; .
\ee
These quantities depend on time through the slow variables $z$ and 
$\al$, because of which one may say that the frequency and attenuation 
experience {\it dynamic shift}. In the case of good resonance, when 
$\al$ is given by Eq. (48), expressions (49) and (50) are
\be
\label{51}
\Om =  \tilde\om_0 = \om_0 -(2S-1)Dz \; , \qquad
\Gm =\gm_2 - \al z \; .
\ee
Also, we define the effective force
\be
\label{52}
f_1 \equiv -i\nu_0 - 2i (\nu_1 +\bt)\cos\om t +\xi \; .
\ee
Then Eqs. (33) to (35) rearrange to
\be
\label{53}
\frac{dx}{dt} = - i(\Om +\xi_0 -i\Gm) x + f_1 z - \al z x^* \; ,
\ee
\be
\label{54}
\frac{dy}{dt} =  - 2\Gm y + \left ( x^* f_1 + f_1^* x\right ) z -
\al z\left [ (x^*)^2 + x^2 \right ] \; ,
\ee
\be
\label{55}
\frac{dz}{dt} = - \al y -\; \frac{1}{2}\left ( x^* f_1 + f_1^* x\right )
- \gm_1 (z-\sgm) +\frac{1}{2}\; \al \left [ (x^*)^2 + x^2 \right ] \; .
\ee
The time derivatives of $\al$ and $\bt$ are proportional to $\gm$,
because of which these functions are to be treated as slow, compared
to $x$. Hence, $\al$ and $\bt$, as well as $\Om$ and $\Gm$, are 
temporal quasi-invariants with respect to $x$.

The solution to Eq. (53), with the quasi-invariants kept fixed, reads
$$
x = x_0 e^{-(i\Om +\Gm)t} \exp\left \{ - i\int_0^t \xi_0(t')\; dt'
\right \} + 
$$
\be
\label{56}
+ z \int_0^t f_1(t') e^{-(i\Om +\Gm)(t-t')}
\exp\left\{ - i \int_{t'}^t \xi_0(t'')\; dt''\right\} \; dt' \; .
\ee
The counterrotating term of Eq. (53), containing $x^*$, gives to the
form (56) a small addition of order $\gm_2/\om_0$, because of which the
latter is omitted.

Define the quantities
\be
\label{57}
\tilde\Gm \equiv \Gm + \gm_3 \; , \qquad \dlt \equiv \om -|\Om| \; ,
\ee
which in the case of good resonance are
$$
\tilde\Gm \approx \gm_2 - \al z + \gm_3 \; , \qquad 
\dlt \approx \om - |\tilde\om_0| =\tilde\Dlt \; .
$$
Averaging Eq. (56) over the stochastic variable $\xi_0$, we get
$$
\ll x \gg \; = -\; \frac{\nu_0 z}{\Om-i\tilde\Gm} +
\frac{(\nu_1+\bt)z}{\dlt+i\tilde\Gm}\; e^{-i\om t} 
+ \left [ x_0 + \frac{\nu_0 z}{\Om-i\tilde\Gm}\; - \; 
\frac{(\nu_1 +\bt)z}{\dlt+i\tilde\Gm} \right ] 
e^{-(i\Om+\tilde\Gm)t} \; .
$$

The fast solution (56) has to be substituted into Eqs. (54) and (55) 
for the slow variables, whose right-hand sides are to be averaged 
over stochastic fields and over time in the infinite interval. Take
the initial condition for the transition function $x$ in the real form
$$
x_0 = \frac{1}{S}\; < S_i^x(0)> \; ,
$$
which is not principal but just slightly simplifies the following 
formulas. Also, keep in mind that the collective attenuation is small,
compared to the collective frequency,
$$
\left | \frac{\Gm}{\Om}\right | \ll 1 \; .
$$
From here, since $\gm_3\ll\om_0$, it follows that $|\tilde\Gm|\ll|\Om|$.
And let us introduce the {\it effective attenuation}
\be
\label{58}
J \equiv \gm_3 - \al\; \frac{\nu_0^2}{\Om^2}\; z - \; 
\frac{\nu_0(\nu_1+\bt)\tilde\Gm}{\Om^2}\; e^{-\tilde\Gm t} +
\frac{(\nu_1+\bt)^2\tilde\Gm}{\dlt^2+\tilde\Gm^2}\left ( 1 -
e^{-\tilde\Gm t}\right ) \; .
\ee
In the latter, the terms related to the action of the constant transverse
field are of second and third order in small parameters. Omitting these 
terms gives
$$
J = \gm_3 + \frac{(\nu_1+\bt)^2\tilde\Gm}{\tilde\Gm^2+\dlt^2}
\left ( 1 - e^{-\tilde\Gm t}\right ) \; .
$$
Averaging the right-hand sides of Eqs. (54) and (55) over the stochastic 
fields and time, we come to the evolution equations

\be
\label{59}
\frac{dy}{dt} = - 2(\gm_2 -\al z) y + 2J z^2 \; ,
\ee
\be
\label{60}
\frac{dz}{dt} = - \al y - J z - \gm_1 (z-\sgm) \; ,
\ee
whose solutions are called the {\it guiding centers}.

As is evident, the derived equations (59) and (60) for the guiding 
centers are incomparably easier to analyse than the initial equations 
(33) to (35). Since Eqs. (59) and (60) are nonlinear, their solutions 
can display rather nontrivial behaviour, especially when the resonant
transverse field or electromotive force are present. Spin dynamics,
in the presence of such transverse injected fields is known to be
quite complicated [13,77,78]. But even when external fields are absent,
the nonlinear spin dynamics may demonstrate very interesting effects.

Before analysing the evolution equations (59) and (60), let us pay 
attention to the physical conditions under which these equations have 
been obtained. The principal point here is the resonance condition (47).
It is only when the spin system is in resonance with the coupled 
electric circuit that we could expect the appearance of noticeable 
transition coherence, which would develop in a self-organized way. This
coherence, of course, can be induced by external transverse fields.
However recall that our major aim is to consider superradiant regimes 
when the transition coherence and, respectively, radiation coherence, 
arises spontaneously, without being stimulated by intensive external
fields.

There can be several possibilities for realizing the resonance 
condition (47). First of all, for systems with spin one-half, there 
is no magnetic anisotropy so that the effective transition frequency 
(31) coincides with the Zeeman frequency, $\tilde\om_0=\om_0$. Then 
we have the same situation as for nuclear magnets [14--20] with $S=1/2$.

If we are dealing with higher spins, then, nevertheless, there is 
the possibility of reducing the problem to an effective spin-one-half 
system. This can be done by tuning the resonant electric circuit to 
one of the transition frequencies of admissible $2S$ transitions and 
by supporting the population of the upper level with the help of a 
permanent nonresonant pumping, as it was accomplished for the nuclear 
spin $I=5/2$ of $^{27}$Al in experiments [11--13]. In that case solely 
the regime of pulsing spin superradiance can be achieved.

For higher spins, the transition from the state with the spin 
projection $S$ to that with the projection $-S$ corresponds to a 
multiphoton transition through $2S-1$ intermediate levels. If the 
longitudinal external magnetic field is sufficiently strong, such 
that $(2S-1)D\ll\om_0$, then again $\tilde\om_0\approx\om_0$, and 
the considered $2S$-photon transition is not much different from the 
one-photon transition, with the same theory [14--20] being applicable.

The worst case occurs when we are interested in a multiphoton 
transition in a spin sample with high magnetic anisotropy, such that 
$(2S-1)D\gg\om_0$. Then the effective transition frequency (31) is 
$\tilde\om_0\simeq -(2S-1)Dz$, which changes with time together with 
$z$. Because of this, to keep the resonance condition (47) valid, one
should respectively vary with time the natural resonator frequency 
$\om$. It is feasible, in principle, to imagine such a {\it sliding  
resonance}, when the circuit characteristics, as inductance and 
capacity, are changing in time so that to preserve the approximate 
equality $\om\approx\tilde\om_0$, following the varying $\tilde\om_0$. 
But it looks that such a sliding resonance would be difficult to realize
experimentally. However, even not this is the major obstacle. When the
magnetic anisotropy is the prevailing part in the effective frequency
(31), so that $\tilde\om_0\simeq -(2S-1)Dz$, then the spin-resonator 
coupling (48) is proportional to $-Dz$. As a result, $\al z\sim -Dz^2$,
which is always negative for the easy-axis anisotropy, with $D>0$. Hence,
the collective attenuation $\Gm$, given in Eq. (51), is always positive.
This means, according to Eqs. (53), (54), or (59), that the function 
$y$ decreases with time. Therefore, there is no generation of coherent 
radiation. For this to occur, the collective frequency $\Gm$ must, at 
least for some period of time, become negative. As is evident from Eqs. 
(54) and (59), a negative attenuation $\Gm$ leads to the generation 
of coherent radiation, but a positive $\Gm$ leads  to the decay of 
transition coherence. Thus, a too strong easy-axis magnetic anisotropy
suppresses transverse coherence, hindering the development of multiphoton
spin superradiance.

In the case of an easy-plane anisotropy, with $D<0$, the quantity
$\al z\sim|D|z^2$ is positive. Hence, $\Gm=\gm_2-\al z$ could be negative.
However, a strong $x-y$-plane anisotropy implies that it is difficult 
to noticeably polarize spins in the $z$-direction. That is, $z$ is always
small, which prevents making $\al z$ large, such that $\al z>\gm_2$.
Therefore a strong single-site anisotropy is an obstacle for achieving
a well-developed transition coherence.

\section{Quantum Stage}

Let us assume that there are no strong external fields imposing on the 
spin system an essential transverse coherence, so that at the initial
stage the motion of transverse spins is not coherent. This is the most
interesting case to consider how the transverse coherence develops in 
a self-organized way from the initially incoherent motion. The period 
of time, when there are yet no collective correlations but only quantum
spin interactions are present can be called the {\it quantum stage}.

At the very beginning of the process, when $\gm t\ll 1$, the 
spin-resonator coupling function (48) is yet close to zero, and we may 
set $\al\ra 0$. If there are no transverse external resonator fields,
which implies that $\nu_1=\nu_c=0$, hence $\bt=0$, then the effective 
attenuation (58) reduces to $J=\gm_3$. In that case, when collective
effects, due to the correlation of spins through the feedback field
have not yet been developed, the evolution equations (59) and (60) are
yet quite simple, having the form
\be
\label{61}
\frac{dy}{dt} = -2\gm_2 y + 2\gm_3 z^2 \; , \qquad
\frac{dz}{dt} = -(\gm_1 +\gm_3) z + \gm_1 \sgm \; .
\ee
Then at short time $t\ra 0$, we have
\be
\label{62}
y \simeq \left ( y_0 -\; \frac{\gm_3 z_0^2}{\gm_2}\right ) e^{-2\gm_2 t}
+ \frac{\gm_3 z_0^2}{\gm_2} \; , \qquad
z \simeq \left ( z_0 -\; \frac{\gm_1\sgm}{\gm_1+\gm_3}\right ) 
e^{-(\gm_1+\gm_3) t} + \frac{\gm_1\sgm}{\gm_1+\gm_3} \; .
\ee

With time, spin correlations increase owing to the resonator feedback 
field. Strengthening spin correlations result in the developing 
transverse coherence. When the transition coherence is well developed,
collective phenomena come into play and the dynamic behaviour of spins
becomes qualitatively different from that at the initial stage. The 
change in the features of spin motion occurs, of course, gradually.
However, it is possible to define the moment of time separating these 
two regimes of motion. The qualitative change in spin dynamics occurs
when the collective attenuation width (50) changes its sign because
of the growing spin-resonator coupling. Then the quantum stage transfers
to the coherent stage. The {\it crossover time} $t_c$, separating these
stages is given by the equation
\be
\label{63}
\Gm(t_c) = 0 \; .
\ee
The latter, in the case of good resonance, takes the form
\be
\label{64}
\al(t_c)\; z(t_c) = \gm_2 \; .
\ee
Introducing the effective spin-resonator coupling parameter
\be
\label{65}
g\equiv \frac{\gm\gm_0\tilde\om_0}{\gm_2(\gm^2 +\tilde\Dlt^2)} \; ,
\ee
in which
\be
\label{66}
\tilde\om_0 \equiv \om_0 - (2S-1) D z_0 \; , \qquad
\tilde\Dlt \equiv \om -|\tilde\om_0| \; ,
\ee
we obtain from Eq. (64) the {\it crossover time}
\be
\label{67}
t_c = \tau \ln \left ( \frac{gz_0}{gz_0 -1} \right ) \; .
\ee
The latter, to be positive and finite, requires that
\be
\label{68}
gz_0 > 1 \qquad ( 0< t_c < \infty) \; .
\ee
When condition (68) is not valid, then the quantum stage will be never 
replaced by the coherent one. For instance, if $gz_0 =1$, then 
$t_c\ra\infty$. When the initial spin polarization is sufficiently high 
and the spin-resonator coupling (65) is strong, then the crossover time 
(67) is
$$
t_c \simeq \frac{\tau}{gz_0} \qquad (gz_0 \gg 1) \; .
$$

For the coherent regime to appear, it is necessary that the crossover 
time (67) be shorter than the following relaxation times,
\be
\label{69}
t_c \ll \left\{ T_1\equiv \frac{1}{\gm_1} \; , \; 
T_2\equiv\frac{1}{\gm_2}\; , \; T_3\equiv \frac{1}{\gm_3}\right\} \; .
\ee
In the other case, spin polarization would relax before the transition 
coherence could arise. Then the solutions (62) at the crossover time 
can be written as
\be
\label{70}
y(t_c)\simeq y_0 + 2\gm_3 t_c z_0^2 \; , \qquad
z(t_c) \simeq z_0 + \gm_1 t_c \sgm \; .
\ee

If the single-site anisotropy is strong, such that $(2S-1)Dz_0\gg\om_0$,
then $\tilde\om_0\sim -z_0$. Since the spin-resonator coupling (65) is
proportional to $\tilde\om_0$, then $g\sim -z_0$, from where it follows 
that $gz_0\sim -z_0^2$, which is always nonpositive. Hence, condition 
(68) cannot be hod true. This means that the coherent stage can never 
appear. In order that the transition coherence could develop, the external
magnetic field $B_0$ has to be sufficiently strong so that to suppress 
the destructive role of the magnetic anisotropy.

The dynamic broadening width $\gm_3$ in Eqs. (61) has been assumed to be 
a constant. In general, it can be a function of time. For example,
considering the dynamics of the spin polarization at the beginning  of
the process, when $t\ra 0$, and when $\gm_1\ll\gm_3$, we have the equation
$$
\frac{dz}{dt} = -\gm_3 z \qquad (t\ra 0) \; .
$$
Setting here
$$
\gm_3(z) \simeq \frac{\gm_s z_0}{2(z_0-z)} \qquad (z\ra z_0)
$$
results in the square-root relaxation [66]
$$
z \simeq z_0(1-\sqrt{\gm_s t})
$$
observed in molecular magnets [45] at short times, below the blocking 
temperature. Thus, for Mn$_{12}$ one has $\gm_s\approx 10^{-2}$ s.

At longer times, below the blocking temperature, the relaxation of
magnetization in molecular magnets follows a stretched exponential law
[37,42,44,49,51,52]. This can be derived if
$$
\gm_3(t) = \frac{\kappa}{t}\; (\gm_* t)^\kappa \qquad
(0<\kappa\leq 1) \; ,
$$
which gives
$$
z=z_0\exp\left\{ -(\gm_* t)^\kappa\right\} \; .
$$
The power $\kappa\approx 0.5$ for $T<T_B$ and $\kappa=1$ for $T>T_B$. The
relaxation parameter $\gm_*$ depends on the applied magnetic field.

Recall that all the consideration of this section concerns the situation
when collective effects have not yet come into play, so that the 
spin-resonator coupling is negligible, $\al\approx 0$. But if the spin 
system is coupled to a resonator and condition (68) is valid, then there 
exists a finite crossover time (67) when the quantum stage is followed by 
the coherent one.

\section{Coherent Stage}

After the crossover time (67), the coupling function (48) fastly 
increases to the value $\al\approx g\gm_2$. This means that collective 
effects come into play and the transition coherence starts developing.
At the transient stage, when time is larger than $t_c$ but much shorter
than $T_1$, one may neglect the longitudinal relaxation, omitting $\gm_1$
in Eq. (60). Superradiance occurring at this transient stage, when
\be
\label{71}
t_c\leq t \ll T_1 \; ,
\ee
is the {\it transient superradiance}.

If there are no transverse external fields, then the effective attenuation
(58) is $J=\gm_3$. For a sufficiently large coupling parameter (65), such 
that $g\gm_2\gg \gm_3$, one may also neglect in Eqs. (59) and (60) the 
terms containing $\gm_3$. Then these equations read
\be
\label{72}
\frac{dy}{dt} = -2\gm_2(1-gz) y \; , \qquad
\frac{dz}{dt} = - g\gm_2 y \; .
\ee
Equations (72) can be solved exactly [15--18] yielding
\be
\label{73}
y =\left (\frac{\gm_p}{g\gm_2}\right )^2 {\rm sech}^2 \left (
\frac{t-t_0}{\tau_p} \right ) \; , \qquad
z= -\; \frac{\gm_p}{g\gm_2}{\rm tanh}\left ( \frac{t-t_0}{\tau_p}
\right ) + \frac{1}{g} \; ,
\ee
where the pulse width $\gm_p$, the pulse time $\tau_p$, with the relation
$\gm_p\tau_p\equiv 1$, and the delay time $t_0$ are the integration 
constants to be defined from the initial conditions taken at the crossover
time $t_c$. Equating the functions (73) to the values (70) gives for the
pulse width the equation
\be
\label{74}
\gm_p^2  =\gm_g^2 +(g\gm_2)^2 (y_0 + 2\gm_3 t_c z_0^2) \; , \qquad
\gm_g\equiv \gm_2(1-gz_0) \; , \qquad \gm_p\tau_p\equiv 1 \; ,
\ee
and for the delay time, we get
\be
\label{75}
t_0 = t_c + \frac{\tau_p}{2}\ln\left | \frac{\gm_p -\gm_g}{\gm_p+\gm_g}
\right | \; .
\ee

Solutions (73) describe a superradiant pulse. Since, by definition, 
superradiance is a self-organized process, we consider the case when 
there is no strong transverse coherence imposed on the system at the 
initial time. This implies the inequality
\be
\label{76}
g^2 y_0 < 1 \; .
\ee
In order that the delay time (75) would be finite, and larger than the 
crossover time, so that
\be
\label{77}
t_c < t_0 < \infty \; ,
\ee
it is necessary that
\be
\label{78}
y_0 + 2\gm_3 t_c z_0^2 > 0 \; , \qquad gz_0 > 1 \; .
\ee
One more condition on superradiance is that the pulse time be shorter
than the spin-spin dephasing time,
\be
\label{79}
\tau_p < T_2 \; .
\ee
Then, taking into account that $\gm_3 t_c\ll 1$, we may distinguish 
two superradiant regimes: {\it triggered superradiance}, when
\be
\label{80}
gz_0 > 1 +\sqrt{1-g^2y_0} \; , \qquad y_0 \neq 0 \; ,
\ee
and {\it pure superradiance}, when
\be
\label{81}
gz_0 > 2 \; , \qquad y_0 = 0 \; .
\ee

One of the necessary conditions for superradiance to occur is $gz_0>1$, 
which is the same condition (68) for the existence of the crossover time
(67). This condition, in view of Eqs. (65) and (66), can be written as
$$
\om_0 z_0 > (2S-1)Dz_0^2 + \frac{\gm_2(\gm^2+\tilde\Dlt^2)}{\gm\gm_0}\; ,
$$
which tells us that the longitudinal magnetic field $B_0$ has to be 
sufficiently strong and the initial spin polarization must be positive, 
$z_0>0$.

For the pure spin superradiance, when $y_0=0$, taking into consideration 
that $\gm_3 t_c \ll 1$, the pulse width and time can be presented as
\be
\label{82}
\gm_p = (gz_0 -1) \gm_2 (1 + \gm_3\tilde t_c) \; , \qquad
\tau_p =\frac{T_2}{gz_0 -1} \;(1 -\gm_3\tilde t_c) \; ,
\ee
and the delay time (75) takes the form
\be
\label{83}
t_0 = t_c + \frac{\tau_p}{2}\; \ln\left | \frac{2}{\gm_3\tilde t_c}
\right | \; ,
\ee
where the notation
$$
\tilde t_c \equiv t_c e^{2\gm t_c} = 
\tau\left ( \frac{gz_0}{gz_0-1}\right )^2 \ln\left (
\frac{gz_0}{gz_0-1}\right )
$$
is used. When $\gm t_c\ll 1$, which is equivalent to the inequality
$gz_0\gg 1$, then $\tilde t_c\simeq t_c$.

At the delay time $t=t_0$, the coherence intensity $y$ is maximal, which
follows from the solutions (73) giving
\be
\label{84}
y(t_0) =\left ( z_0 - \; \frac{1}{g}\right )^2 ( 1 + 2\gm_3 t_c) \; ,
\qquad z(t_0) =\frac{1}{g} \; .
\ee
After this, the transition coherence fastly decays. Thus, for $t\gg t_0$,
one has
\be
\label{85}
y\simeq 4y(t_0) e^{-2\gm_p t} \; , \qquad
z\simeq - z_0 + \frac{2}{g} \; .
\ee
For large $g\gg 1$, the initial spin polarization becomes almost 
completely inverted. This effect can be used for fast polarization 
reversal in polarized solid-state targets employed in scattering 
experiments [10,79]. The polarization reversal under a superradiant 
burst is illustrated in Fig. 1, which results from numerical 
simulations [21--23].

\section{Pulsing Regime}

After the transient superradiant burst, occurring at the delay time 
$t_0$, the transition coherence dies out, unless the inversion of spin 
polarization is supported by a permanent pumping. This pumping can be 
accomplished by means of dynamic nuclear polarization. In the latter 
case, the regime of pulsing spin superradiance can be achieved. To 
study this regime, we consider a spin system without external transverse 
fields. Then the effective attenuation (58) is $J=\gm_3$. At long times,
such that $\gm t\gg 1$, the coupling function (48) reaches its maximal 
value $\al\cong g\gm_2$. In that case, Eqs. (59) and (60) can be written 
as
\be
\label{86}
\frac{dy}{dt} =  v_1 \; , \qquad \frac{dz}{dt} = v_2 \; ,
\ee
with the right-hand sides
\be
\label{87}
v_1 = -2\gm_2(1-gz)y + 2\gm_3z^2 \; , \qquad
v_2 = -g\gm_2 y - \gm_3 z - \gm_1^*(z-\sgm) \; .
\ee
In the presence of the permanent polarization pumping, the pumping rate 
$\gm_1^*$ takes the place of the spin-lattice relaxation parameter 
$\gm_1$, while $\sgm\in[-1,1]$ is a pumping parameter.

To understand the behaviour of the solutions $y$ and $z$ at long times
$t\ra\infty$, let us find the stationary points of Eqs. (86), which are 
given by the equations $v_1=v_2=0$. In general, these equations give us
two stationary points $y_1^*,\; z_1^*$ and $y_2^*,\; z_2^*$, also called
fixed points. Among these, however, only those have sense, which satisfy 
the physical restrictions
\be
\label{88}
0\leq y \leq 1 \; , \qquad -1\leq z \leq 1 \; .
\ee
Also, the actual solutions of Eq. (86) will tend to that fixed point
which is stable, whose properties can be derived by means of the 
Lyapunov stability analysis. The stability of fixed points depends on 
the values of the spin-resonator coupling (65) and of the effective 
pumping parameter
\be
\label{89}
\sgm^* \equiv \frac{\gm_1^*\sgm}{\gm_1^* + \gm_3} \; .
\ee
The value of this parameter depends on the relation between the pumping 
rate $\gm_1^*$ and the dynamic broadening width $\gm_3$. Since the 
effects of temperature can be modelled through a thermal bath of random 
fluctuations [80], the temperature dependence can be incorporated in 
$\gm_3$.

When the pumping is weak, such that
\be
\label{90}
g\sgm^* \leq -1 \; ,
\ee
then the stable fixed point corresponds to the solutions
\be
\label{91}
y_1^* \simeq \frac{(\gm_1^*+\gm_3)\gm_3}{g^2\gm_1^*\gm_2}\; 
\left | 1+ g\sgm^*\right | \; , \qquad
z_1^* \simeq \sgm\left ( 1 + \frac{\gm_3}{g\sgm\gm_1^*}\right ) \; .
\ee
The related characteristic exponents are
\be
\label{92}
X_1^+ \simeq -\gm_1^* - \; \frac{\gm_1^* -2\gm_2}{g\sgm\gm_2}\; \gm_3\; ,
\qquad X_1^- \simeq - 2\gm_2 (1 -g\sgm) -\gm_3 \; .
\ee
The fixed point (91) is a stable node. The coherence intensity 
$y_1^*\sim \gm_3/g\gm_2$ is small.

When the spin-resonator coupling is weak, so that
\be
\label{93}
|g\sgm^*| \ll 1 \; ,
\ee
then the stationary point is given by the solutions
\be
\label{94}
y_1^* \simeq \frac{\gm_1^*\gm_3(\sgm^*)^2}{(\gm_1^*+\gm_3)\gm_2}
\left ( 1 + \frac{\gm_1^* -\gm_3}{\gm_1^*+\gm_3} \; g\sgm^*\right )\; ,
\qquad z_1^* \simeq \sgm^*\left ( 1 - \;
\frac{\gm_3}{\gm_1^*+\gm_3}\; g\sgm^*\right ) \; .
\ee
The corresponding characteristic exponents
\be
\label{95}
X_1^+ \simeq -\gm_1^* - \gm_3\left ( 1  - \;
\frac{4\gm_2}{\gm_1^* -2\gm_2+\gm_3}\; g\sgm^*\right ) \; , \qquad
X_1^- \simeq - 2\gm_2 \left ( 1  -\; 
\frac{\gm_1^* -2\gm_2 -\gm_3}{\gm_1^*-2\gm_2+\gm_3}\; g\sgm^*\right )
\ee
show that the fixed point (94) is, as early, a stable node. The limiting 
value $y_1^*\ll 1$. In these both cases of either Eq. (90) or Eq. (93),
the solutions monotonically tend to the stationary point $y_1^*,\; z_1^*$.

When the spin-resonator coupling as well as pumping are sufficiently
strong, so that
\be
\label{96}
g\sgm^*\geq 1 \; ,
\ee
then the solutions of the evolution equations (86) tend to another fixed
point
\be
\label{97}
y_2^* \simeq \frac{\gm_1^*+\gm_3}{g^2\gm_2}\left ( g\sgm^* -1\right )\; ,
\qquad z_2^* \simeq \frac{1}{g}\left ( 1  -\;
\frac{\gm_3}{g\sgm\gm_1^*}\right ) \; .
\ee
Now, the characteristic exponents are complex valued,
\be
\label{98}
X_2^\pm \simeq -\; \frac{1}{2}\; \gm_1^* - \; \frac{1}{2}\; \gm_3
\left ( 1 + \frac{2\gm_2}{g\sgm\gm_1^*}\right ) \pm i\om_\infty \; ,
\ee
containing the asymptotic frequency
\be
\label{99}
\om_\infty \simeq \sqrt{2g\sgm\gm_1^*\gm_2} \; .
\ee
The fixed point (97) is a stable focus. The limiting value of the 
coherence intensity $y_2^*\sim \gm_1^*/g\gm_2$ is small. However, in
the way to the fixed point (97), the spin system exhibits a series of
superradiant bursts, which can be demonstrated by solving Eqs. (86) 
numerically [20,30,71]. This is the regime of {\it pulsing spin 
superradiance}. The superradiant bursts are not equidistant in time,
and become approximately periodic only at $t\ra\infty$, with the 
asymptotic period
\be
\label{100}
T_\infty \equiv \frac{2\pi}{\om_\infty} \simeq 
\pi\sqrt{\frac{2T_1^*T_2}{g\sgm}}\; ,
\ee
where $\gm_1^*T_1^*\equiv 1$. Each superradiant burst, occurring in the
time interval $0<t<T_1^*$, possesses a high level of transition coherence, 
but after $T_1^*$ the amplitude of the coherence intensity $y$ diminishes,
tending to $y_2^*$.

\vskip 3mm

In conclusion, superradiant regimes may be employed in the operation 
of spin masers. There are three possible such regimes: {\it triggered 
superradiance}, {\it pure superradiance}, and {\it pulsing superradiance}.
In materials with higher spins, as molecular magnets, it is necessary to
take into account single-site magnetic anisotropy. If this anisotropy
is strong, it hinders the possibility of achieving a high level of 
transition coherence in multiphoton transitions. However, by a sufficiently
strong external magnetic field the destructive role of magnetic anisotropy
can be suppressed. Another possibility is to tune the resonant electric
circuit to one of the quantum transition lines of a high spin. When a sole 
transition line is selected by the resonant tuning, the situation becomes
similar to the case of an effective two-level system

\vskip 5mm

{\bf Acknowledgement}

\vskip 3mm

I devote this work to my friend Sven Hartmann, who is one of the first
persons with whom I have discussed the intricate problems of spin 
superradiance, benefiting so much from these discussions. I have started 
this paper when visiting the Ames laboratory at the Iowa State University, where I had a 
fortunate possibility of enjoying many fruitful discussions with 
B.N. Harmon, whom I am very grateful for useful advice and hospitality.
I also very much appreciate helpful discussions with F. Borsa, 
V.V. Dobrovitskii, B.C. Gerstein, J.T. Manassah, and M. Pruski.

\vskip 2mm

Financial support from the Iowa State University, USA, and from the
S\~ao Paulo State Research Foundation, Brazil, is acknowledged.

\newpage

\begin{center}
{\bf Figure Caption}
\end{center}

{\bf Fig. 1}. Intensity of transition coherence $y$ in arbitrary units 
(upper curve) and the longitudinal spin polarization $z$ (lower curve) 
as functions of time measured in units of $T_2$, obtained from numerical
simulations for 300 spins.


\newpage

\begin{thebibliography}{99}

\bibitem{1}
Seigman, A.E., 1964, {\it Microwave Solid-State Masers} (New York:
McGraw-Hill).

\bibitem{2}
Friedberg, R. and Hartmann, S.R., 1974, {\it Phys. Rev. B}, {\bf 10}, 
1728.

\bibitem{3}
Romalis, M.V. and Happer, W., 1999, {\it Phys. Rev. A}, {\bf 60}, 
1385.

\bibitem{4}
Andreev, A.V., Emelyanov, V.I., and Ilinski, Y.A., 1993, 
{\it Cooperative Effects in Optics} (Bristol: Institute of Physics).

\bibitem{5}
Benedict, M.G., Ermolaev, A.M., Malushev, V.A., Sokolov, I.V., and 
Trifonov, E.D., 1996, {\it Superradiance-Multiatomic Coherent Emission}
(Bristol: Institute of Physics).

\bibitem{6}
Bloembergen, N. and Pound, R.V., 1954, {\it Phys. Rev.}, {\bf 95}, 8.

\bibitem{7}
Kiselev, J.F., Prudkoglyad, A.F., Shumovsky, A.S., and Yukalov, V.I., 
1988, {\it Mod. Phys. Lett. B}, {\bf 1}, 409.

\bibitem{8}
Kiselev, Y.F., Prudkoglyad, A.F., Shumovsky, A.S., and Yukalov, V.I., 
1988, {\it J. Exp. Theor. Phys.}, {\bf 67}, 413.

\bibitem{9}
Bazhanov, N.A. et al., 1990, {\it J. Exp. Theor. Phys.}, {\bf 70}, 1128.

\bibitem{10}
Reichertz, L.A. et al., 1994, {\it Nucl. Instrum. Methods Phys.
Res. A}, {\bf 340}, 278.

\bibitem{11}
B\"osiger, P., Brun, E., and Meier, D., 1978, {\it Phys. Rev. A}, 
{\bf 18}, 671.

\bibitem{12}
B\"osiger, P., Brun, E., and Meier, D., 1979, {\it Phys. Rev. A}, 
{\bf 20}, 1073.

\bibitem{13}
Holzner, R., Derighetti, B., Ravani, M., and Brun, E., 1987, {\it Phys. 
Rev. A}, {\bf 36}, 1280.

\bibitem{14}
Yukalov, V.I., 1993, {\it Laser Phys.}, {\bf 3}, 870.

\bibitem{15}
Yukalov, V.I., 1995, {\it Phys. Rev. Lett.}, {\bf 75}, 3000.

\bibitem{16}
Yukalov, V.I., 1995, {\it Laser Phys.}, {\bf 5}, 526.

\bibitem{17}
Yukalov, V.I., 1995, {\it Laser Phys.}, {\bf 5}, 970.

\bibitem{18}
Yukalov, V.I., 1996, {\it Phys. Rev. B}, {\bf 53}, 9232.

\bibitem{19}
Yukalov, V.I., 1997, {\it Laser Phys.}, {\bf 7}, 58.

\bibitem{20}
Yukalov, V.I. and Yukalova, E.P., 1998, {\it Laser Phys.}, {\bf 8}, 1029.

\bibitem{21}
Belozerova, T.S., Henner, V.K., and Yukalov, V.I., 1992, {\it Phys. 
Rev. B}, {\bf 46}, 682.

\bibitem{22}
Belozerova, T.S., Henner, V.K., and Yukalov, V.I., 1992, {\it Laser 
Phys.}, {\bf 2}, 545.

\bibitem{23}
Belozerova, T.S., Henner, V.K., and Yukalov, V.I., 1992, {\it Comput. 
Phys. Commun.}, {\bf 73}, 151.

\bibitem{24}
Belozerova, T.S., Henner, V.K., and Yukalov, V.I., 1992, {\it Tech. 
Phys. Lett.}, {\bf 18}, 404.

\bibitem{25}
Belozerova, T.S., Henner, V.K., and Yukalov, V.I., 1994, {\it Proc. 
Int. Soc. Opt. Eng.}, {\bf 2098}, 86.

\bibitem{26}
Belozerova, T.S., Davis, C.L., and Henner, V.K., 1998, {\it  
Phys. Rev. B}, {\bf 58}, 3111.

\bibitem{27}
Belozerova, T.S., Davis, C.L., and Henner, V.K., 1999, {\it  
Comput. Phys. Commun.}, {\bf 121}, 214.

\bibitem{28}
Yukalov, V.I., Cottam, M.G., and Singh, M.R., 1999, {\it Phys. Rev. B},
{\bf 60}, 1227.

\bibitem{29}
Yukalov, V.I., Cottam, M.G., and Singh, M.R., 1999, {\it J. Appl. Phys.},
{\bf 85}, 5627.

\bibitem{30}
Yukalov, V.I. and Yukalova, E.P., 2001, {\it Laser Phys.}, {\bf 11}, 546.

\bibitem{31}
Yukalov, V.I., 1992, {\it Laser Phys.}, {\bf 2}, 559.

\bibitem{32}
Yukalov, V.I., 1997, {\it Proc. Int. Soc. Opt. Eng.}, {\bf 3239}, 118.

\bibitem{33}
Kodama, R.H., 1999, {\it J. Magn. Magn. Mat.}, {\bf 200}, 359.

\bibitem{34}
Hadjipanays, G.C., 1999, {\it J. Magn. Magn. Mat.}, {\bf 200}, 373.

\bibitem{35}
Kahn, O., 1993, {\it Molecular Magnetism} (New York: VCH).

\bibitem{36}
Barbara, B. {\it et al.}, 1999, {\it J. Magn. Magn. Mat.}, {\bf 200}, 167.
 
\bibitem{37}
Caneschi, A. {\it et al.}, 1999, {\it J. Magn. Magn. Mat.}, {\bf 200}, 
182.

\bibitem{38}
Hartmann-Bourton, F., Politi, P., and Villain, J., 1996, {\it Int. J. 
Mod. Phys. B}, {\bf 10}, 2577.

\bibitem{39}
Garanin, D.A., and Chudnovsky, E.M., 1997, {\it Phys. Rev. B}, {\bf 56}, 
11102.

\bibitem{40}
Lascialfari, A. {\it et al.}, 1998, {\it Phys. Rev. Lett.}, {\bf 81}, 
3773.

\bibitem{41}
Lascialfari, A. {\it et al.}, 1998, {\it Phys. Rev. B}, {\bf 57}, 
514.

\bibitem{42}
Thomas, L. and Barbara, B., 1998, {\it J. Low Temp. Phys.}, 
{\bf 113}, 1055.

\bibitem{43}
Jang, Z.H., Lascialfari, A., Borsa, F., and Gatteschi, D., 2000, {\it
Phys. Rev. Lett.}, {\bf 84}, 2977.

\bibitem{44}
Bokacheva, L., Kent, A.D., and Walters, M.A., 2000, {\it Phys. Rev. 
Lett.}, {\bf 85}, 4803.

\bibitem{45}
Chiorescu, I. {\it et al.}, 2000, {\it Phys. Rev. Lett.}, {\bf 85}, 4807.

\bibitem{46}
Garanin, D.A., Chudnovsky, E.M., and Schilling, R., 2000, {\it Phys. Rev.
B}, {\bf 61}, 12204.

\bibitem{47}
Zhong, Y., Sarachik, M.P., Yoo, J., and Hendrikson, D.N., 2000, {\it Phys.
Rev. B}, {\bf 62}, 9256.

\bibitem{48}
Furukawa, Y. {\it et al.}, 2000, {\it Phys. Rev. B}, {\bf 62}, 14246.

\bibitem{49}
Sangregorio, C. {\it et al.}, 1997, {\it Phys. Rev. Lett.}, {\bf 78}, 
4645.

\bibitem{50}
Cacuiffo, R. {\it et al.}, 1998, {\it Phys. Rev. Lett.}, {\bf 81}, 4744.

\bibitem{51}
Ohm, T., Sangregorio, C., and Paulsen, C., 1998, {\it Eur. Phys. J. B}, 
{\bf 6}, 195.

\bibitem{52}
Cuccoli, A. {\it et al.}, 1999, {\it Eur. Phys. J. B}, {\bf 12}, 39.

\bibitem{53}
Villain, J. and Fort, A., 2000, {\it Eur. Phys. J. B}, {\bf 17}, 69.

\bibitem{54}
Furukawa, Y., Lascialfari, A., Jang, Z.H., and Borsa, F., 2000, {\it J.
Appl. Phys.}, {\bf 87}, 6265.

\bibitem{55}
Gatteschi, D. {\it at al.}, 1991, {\it Nature}, {\bf 354}, 465.

\bibitem{56}
Dobrovitskii, V.V., Katsnelson, M.I., and Harmon, B.N., 2000, {\it Phys.
Rev. Lett.}, {\bf 84}, 3458.

\bibitem{57}
Lascialfari, A., Gatteschi, D., Borsa, F., and Cornia, A., 1997, 
{\it Phys. Rev. B}, {\bf 55}, 14341.

\bibitem{58}
Julien, M.N. {\it et al.}, 1999, {\it Phys. Rev. Lett.}, {\bf 83}, 227.

\bibitem{59}
Lascialfari, A. {\it et al.}, 2000, {\it Phys. Rev. B}, {\bf 61}, 6839.

\bibitem{60}
Zvezdin, A.K., Dobrovitskii V.V., Harmon, B.N., and Katsnelson, M.I.,
1998, {\it Phys. Rev. B}, {\bf 58}, 14733.

\bibitem{61}
Dobrovitskii, V.V. and Harmon, B.N., 1998, {\it J. Appl. Phys.}, 
{\bf 83}, 6599.

\bibitem{62}
Katsnelson, M.I., Dobrovitskii, V.V., and Harmon, B.N., 1999, {\it
Phys. Rev. B}, {\bf 59}, 6919.

\bibitem{63}
Prokofev, N.V. and Stamp, P.C.E., 1996, {\it J. Low Temp. Phys.}, 
{\bf 104}, 143.

\bibitem{64}
Dobrovitskii, V.V. and Zvezdin, A.K., 1997, {\it Europhys. Lett.}, 
{\bf 38}, 377.

\bibitem{65}
Gunther, L., 1997, {\it Europhys. Lett.}, {\bf 39}, 1.

\bibitem{66}
Prokofev, N.V. and Stamp, P.C.E., 1998 {\it Phys. Rev. Lett.},
{\bf 80}, 5794.

\bibitem{67}
Coleman, A.J. and Yukalov, V.I., 2000, {\it Reduced Density Matrices}
(Berlin: Springer).

\bibitem{68}
Slichter, C.P., 1978, {\it Principles of Magnetic Resonance} 
(Berlin: Springer).

\bibitem{69}
Ernst, R.R., Bodenhausen, G., and Wokaun, A., 1987, {\it Principles of
Nuclear Magnetic Resonance in One and Two Dimensions} (Oxford: Clarendon).

\bibitem{70}
Yukalov, V.I., 1998, {\it Statistical Green's Functions} (Kingston:
Queen's University).

\bibitem{71}
Yukalov, V.I. and Yukalova, E.P., 2000, {\it Phys. Part. Nucl.}, 
{\bf 31}, 1128.

\bibitem{72}
Abragam, A., 1961, {\it Principles of Nuclear Magnetism} (Oxford:
Clarendon).

\bibitem{73}
Tahir-Kheli, R.A. and Ter Haar, D., 1962, {\it Phys. Rev.}, {\bf 127},
88.

\bibitem{74}
Anderson, F.B. and Callen, H.B., 1964, {\it Phys. Rev. A}, {\bf 136}, 
1068.

\bibitem{75}
Lines, M.E., 1967, {\it Phys. Rev.}, {\bf 156}, 534.

\bibitem{76}
Cottam, M.G. and Lockwood, D.J., 1986, {\it Light Scattering in 
Magnetic Solids} (New York: Wiley).

\bibitem{77}
Rukhlov, V.S., 1991, {\it Phys. Lett. A}, {\bf 160}, 131.

\bibitem{78}
Yukalov, V.I., Gonzalez, J.A., and Dias, C.L., 1998, {\it Laser 
Phys.}, {\bf 8}, 19.

\bibitem{79}
Yukalov, V.I., 1996, {\it Nucl. Instrum. Methods Phys. Res. A}, 
{\bf 370}, 345.

\bibitem{80}
Antropov, V.P. and Harmon, B.N., 1996, {\it J. Appl. Phys.}, 
{\bf 79}, 5409.

\end{thebibliography}
\end{document}